# Measurement of lateral and interfacial thermal conductivity of single- and bi-layer $MoS_2$ and $MoSe_2$ using refined optothermal Raman technique

Xian Zhang[1], Dezheng Sun[2], Yilei Li[2], Gwan-Hyoung Lee[3], Xu Cui[1], Daniel Chenet[1], Yumeng You[4], Tony F. Heinz[2], and James C. Hone[1]*

[1]Department of Mechanical Engineering, [4]Departments of Physics, Columbia University, New York, NY, 10027, USA

[2]Department of Applied Physics, Stanford University, Stanford, CA, 94305, USA

[3]Department of Materials Science and Engineering, Yonsei University, Seoul 120-749, Korea

*: corresponding author, jh2228@columbia.edu.

# ABSTRACT

Atomically thin materials such as graphene and semiconducting transition metal dichalcogenides (TMDCs) have attracted extensive interest in recent years, motivating investigation into multiple properties. In this work, we demonstrate a refined version of the optothermal Raman technique[39,40] to measure the thermal transport properties of two TMDC materials, $MoS_2$ and $MoSe_2$, in single-layer (1L) and bi-layer (2L) forms. This new version incorporates two crucial improvements over previous implementations. First, we utilize more direct measurements of the optical absorption of the suspended samples under study and find values ~40% lower than previously assumed. Second, by comparing the response of fully supported and suspended samples using different laser spot sizes, we are able to independently measure the interfacial thermal conductance to the substrate and the lateral thermal conductivity of the supported and suspended materials. The approach is validated by examining the response of a suspended film illuminated in different radial positions. For 1L $MoS_2$ and $MoSe_2$, the room-temperature thermal conductivities are (84±17) W/mK and (59±18) W/mK, respectively. For 2L $MoS_2$ and $MoSe_2$, we obtain values of (77±25) W/mK and (42±13) W/mK. Crucially, the interfacial thermal conductance is found to be of order 0.1-1 $MW/m^2K$, substantially smaller than previously assumed, a finding that has important implications for design and modeling of electronic devices.

Following the interest in graphene since its first isolation by mechanical exfoliation one decade ago,[1-3] the broader family of two-dimensional (2D) materials has become the subject of extensive attention thanks to their unique properties and atomically thin structure.[4-13] In particular, the TMDC materials have shown unique optical and electrical properties, such as band structure transitions,[14-16] semiconducting transport behavior,[8,17] and strong photoluminescence,[8-10,18] which are distinct from those of graphene and other carbon allotropes.[19] TMDC materials are also intriguing for optical, electrical and thermal applications, especially in few-layer forms.[8,10,15,20-32] While electron transport in TMDC materials has been widely studied, there have been only limited experimental data[33,34] published on thermal transport in $MoS_2$ and no experimental data about $MoSe_2$. Moreover, the published work reports lower thermal conductivity for single-layer (1L) forms than the few-layer, which is opposite to the trend discovered on a well-studied 2D material, graphene.[52] Therefore more robust experiments and modeling are needed to explain this trend or to provide more accurate values. Meanwhile, the large disparity in thermal conductivities predicted by theory[35-38] also motivates further experimental investigation. Here we demonstrate an improved experimental method for measuring thermal conductivities of thin TMDCs. This is used to provide new measurements of the thermal conductivity of 1L $MoS_2$, as well as the first measurements for 2L $MoS_2$, 1L $MoSe_2$, and 2L $MoSe_2$. The systematic experimental results enable comparisons of experimental results between different layer numbers, different materials, and also different material conditions (suspended vs. supported). In addition, the method provides a measurement of interfacial thermal conductivity, an important parameter for understanding heat dissipation in electronic devices.

The optothermal Raman technique[39,40] has been the most successful method for measurement of thermal conductivity of 2D materials. In this technique, a laser is focused at the center of a suspended flake and used to measure the position of a Raman-active mode. As the laser power is increased, the sample is heated which enables red-shift Raman mode due to thermal softening. Thermal modeling can then be used to extract the thermal conductivity from the measured shift rate.

The thermal modeling used for determination of thermal conductivity requires additional input of a number of parameters: the rate of mode softening with temperature; optical absorption; and the lateral and interfacial thermal conductance of the supported area of the flake. However, in previous work typically only the mode softening rate is directly measured, while the values of other parameters are derived from published values or assumed. Here we present measurements of the thermal conductivity of $MoS_2$ and $MoSe_2$ in which all of these parameters are determined experimentally. We find values of optical absorption and interfacial conductance that differ substantially from values used previously, and which correspondingly affect the derived thermal conductivity values. We also suspend samples over holes with larger diameter (2.5 - 5 μm vs. 1.2 μm), which helps to minimize effects of the finite spot size. Finally, as a further validation of the model, we measure the response of samples illuminated at different radial positions, and find a position-dependence that fits with the parameters extracted above. We then extend this robust measurement methodology to 2L $MoS_2$, and $MoSe_2$ (1L and 2L), whose thermal conductivity is measured and the thermal conductivities trend of which are studied for the first time. We obtain room-temperature thermal conductivities of (84±17) and (77±25) W/mK for suspended 1L and 2L $MoS_2$, respectively, larger than found in recent published results[34] and comparable to the published result of few-layer $MoS_2$.[33] And our finding of the thermal conductivity of supported 1L $MoS_2$ is in good consistent with the recent publication with an absorption value of ~5%.[54] For 1L and 2L $MoSe_2$, we find values of (59±18) W/mK and (42±13) W/mK. For all the materials, the thermal conductivities are smaller when supported on a substrate, and decrease with increasing temperature, as expected due to phonon-phonon scattering.[51]

## RESULTS AND DISCUSSION

Figure 1a shows a Scanning Electron Microscope (SEM) image of an exfoliated monolayer MoS$_2$ flake transferred onto a substrate with 4μm diameter holes. Photoluminescence (PL) peak intensity mapping (Figure 1a,b) reveals that the area over the hole has peak intensity about 500 times that of the supported area, in accordance with previous reports,[43] which confirms that it is well suspended. Figure 1c shows the $E_{2g}^1$ and $A_{1g}$ Raman peaks for the flake, which are used to determine the number of atomic layers (Figure 1c).[41] Raman and PL methods are also used to determine the atomic layer numbers of MoSe$_2$ flakes (see supplementary materials).[18,44] In addition, the high resolution atomic force microscopy (AFM) method is used as a thickness confirmation. Figure 1d shows a schematic diagram of the experimental setup. The Raman laser was focused on either the center of a supported flake, or on a suspended area of the same flake, using both 50X and 100X objective lenses. A 532nm laser was used for MoS$_2$, and a 633nm laser beam was used for MoSe$_2$. For each objective, the spot size was obtained by scanning across a sharp flake edge, and plotting the measured integrated Raman peak intensity vs. position. By fitting the curves from 5 repeated measurements using a Gaussian error function, we obtained spot sizes of (0.46±0.03) μm and (0.62±0.03) μm for the 100x and 50x objectives, respectively.

We first calibrate the shift rate of the $A_{1g}$ peak position with temperature by heating the entire substrate. Figure 2a shows an example of the temperature-dependent Raman spectra of suspended 1L MoS$_2$. The Raman $A_{1g}$ peak follows a red shift with increasing temperature. The Raman measurements of other samples are all showing the similar trend. Figure 2b is the Raman $A_{1g}$ peak position as a function of temperature on supported and suspended 1L MoS$_2$. The observed linear red shift with increasing frequency can be explained by thermally driven bond softening, as previously observed for graphene.[40] Figures 2b,c,d show the temperature dependence of the $A_{1g}$ peaks for 1L $MoSe_2$, 2L $MoS_2$, and 2L $MoSe_2$. Since $A_{1g}$ is an out-of-plane mode, the smaller red shift for the supported area can be attributed to substrate interactions. Indeed, when the same fit and data analysis was conducted using the in-plane $E_{2g}^1$ mode, no difference was observed between suspended and supported areas. This difference is also absent for 2L samples, which show smaller changes with temperature compared to the 1L samples. This trend agrees with previous work on 1L and multilayer $MoS_2$.[45] Because the $A_{1g}$ mode has higher signal intensity, it was used for thermal transport calculations (discussed in the following paragraphs), but similar results were obtained using the $E_{2g}^1$ mode. The linear shift rate of $A_{1g}$ Raman peak with temperature is defined as the first order temperature coefficient.

For thermal conductivity measurements, the $A_{1g}$ peak shift was measured as a function of laser power for both supported and suspended samples, using 0.46μm and 0.62μm spot sizes. Figure 3a shows the shift rates for 1L $MoS_2$, on both supported and suspended areas. These measurements were repeated for all samples (Figure 3b,c,d). Hole diameters of 4μm, 2.5μm, 3μm, and 3μm were used for 1L $MoS_2$, 1L $MoSe_2$, 2L $MoS_2$, and 2L $MoSe_2$, respectively.

To determine the absolute power absorbed, flakes of each sample type were separately exfoliated onto quartz substrates, and their measured absorption spectra were used to determine the frequency-dependent complex dielectric function of each sample type. This, along with dielectric functions of the substrate materials (gold[46] and $SiO_2$[47]), was then used to calculate the absorbance at 532nm and 633nm using the standard transfer matrix method.[48,53]. For the supported layers, the optical interference effect from the substrate is taken into account in calculating the absorbed power. For the suspended layers, the incident light provides the dominant contribution to the absorption. Because the depth of the holes (1.18 μm), the incident light spreads to a spot size of ~2.5 µm after reflection from the bottom. This diffuse light can account for only 5% of the temperature increase at the center of the suspended flake. We have considered this effect in the thermal conductivity calculations. (More analysis is included in the supplementary material.) The obtained absorbance values, along with the temperature coefficients and power shift rates are summarized in Table 1.

Table 1. First Order Temperature Coefficients, Absorption Coefficients, and Power Shift Rates of Supported and Suspended TMDC Materials

|  | Temperature coefficient (cm$^{-1}$/K) | | Absorption coefficient (%) | | Power shift rate(cm$^{-1}$/µW) | | |
|---|---|---|---|---|---|---|---|
|  | Supported | Suspended | Supported | Suspended | Supported | | Suspended |
|  |  |  |  |  | 0.46µm spot | 0.62µm spot |  |
| 1L MoS$_2$ | 0.0167±0.0007 | 0.0203±0.0006 | 5.2±0.1 | 5.8±0.1 | 0.0204±0.0009 | 0.0112±0.0005 | 0.0987±0.0022 |
| 1L MoSe$_2$ | 0.0111±0.0005 | 0.0141±0.0004 | 5.7±0.1 | 5.6±0.1 | 0.0443±0.0021 | 0.0285±0.0011 | 0.1226±0.0028 |
| 2L MoS$_2$ | 0.0139±0.0003 | 0.0136±0.0006 | 11.5±0.1 | 12.1±0.1 | 0.0108±0.0003 | 0.0057±0.0001 | 0.0323±0.0018 |
| 2L MoSe$_2$ | 0.0095±0.0004 | 0.0094±0.0004 | 9.7±0.1 | 9.4±0.1 | 0.0300±0.0011 | 0.0180±0.0004 | 0.0493±0.0020 |

Figure 1d shows schematic diagrams of heat conduction at steady state when supported and suspended samples are heated by the Raman laser. The radial temperature distribution $T(r)$ is governed by the absorbed power and conduction through the sample to the substrate.[39] Convection through air accounts for less than 0.13% of the total heat conduction and is ignored (see supplementary materials). The total absorbed laser power $P$ is first determined from the laser power and the absorption coefficients (Table 1). Assuming a Gaussian profile and a spot radius of $r_0$, this can then be used to calculate the volumetric heating power density $q'''(r)$:

$$q'''(r) = P \cdot \frac{1}{t} \cdot \frac{1}{\pi r_0^2} \exp\left(-\frac{r^2}{r_0^2}\right) \qquad (1)$$

where $t$ is the thickness of the flake. For a suspended flake, $T(r)$ is then given by

$$\frac{1}{r}\frac{d}{dr}\left(r\frac{dT(r)}{dr}\right) + \frac{q'''(r)}{\kappa} = 0 \quad (0 < r \leq R) \tag{2}$$

$$\frac{1}{r}\frac{d}{dr}\left(r\frac{dT(r)}{dr}\right) - \frac{g}{\kappa_s t}(T(r) - T_a) + \frac{q'''(r)}{\kappa_s} = 0 \quad (r > R) \tag{3}$$

Here $R$ is the hole radius and $T_a$ is the substrate temperature. $\kappa$ and $\kappa_s$ are the thermal conductivity of the suspended and supported portions of the flake, and g is the interfacial thermal conductance between the flake and the substrate. The boundary conditions $(dT)/(dr)|_{r=0}=0$ and $T(r\to\infty)=0$ are also applied. For a supported flake, Eq. 3 is used everywhere.

The measured temperature of the flake center is determined by the measured $A_{1g}$ position using the shift rate determined above (Table 1). This value reflects the local temperature distribution, weighted by the Gaussian profile of the laser spot:

$$T_m = \frac{\int_0^\infty T(r)\exp\left(-\frac{r^2}{r_0^2}\right) r dr}{\int_0^\infty \exp\left(-\frac{r^2}{r_0^2}\right) r dr} \tag{4}$$

The experimental results, as shown in Figure 2,3 and summarized in Table 1, allow calculation of the measured thermal resistance of each sample, given as $R_m = T_m/P$. For the supported 1L MoS$_2$ sample, we find $R_m = (1.222\pm0.074)\times 10^6$ K/W and $(0.671\pm0.041)\times 10^6$ K/W, for the 0.46μm and 0.62μm spot sizes, respectively. To extract values of the fitting parameters $\kappa$, $\kappa_s$, and $g$, we follow the method of Cai, et al.[39] We first examine the case of fully supported samples to obtain the values of $\kappa_s$ and $g$. In brief, solving Eqs. 1-4 yields a value of $R_m$ that is a function of $\kappa_s$ and $g$. Moreover, the ratio of $R_m$ for the two different spot sizes is a function of the ratio $g/\kappa_s$. We therefore use the measured $R_m$ ratio to obtain $g/\kappa_s$, and then use $R_m$ for a single spot size to obtain $\kappa_s$ and $g$ independently. This analysis yields $\kappa_s$= (55±20) W/mK and $g$ = (0.44±0.07) MW/m²K for the supported 1L MoS$_2$. The $\kappa_s$ and $g$ values are obtained by a combination of using 100x and 50x lenses (with spot sizes of 0.46μm and 0.62μm), and confirmed by an additional experiment using 20x lens (with spot size of 1.23μm) which generates the combination of using 50x and 20x lenses. Also with the effect from the bottom of the hole (see supplementary materials), values for the other samples are given in Table 2. The interfacial thermal conductance of MoSe$_2$ is lower because these samples are deposited directly on the SiO$_2$ without a gold film.

Table 2. Room-Temperature Thermal Conductivities and Interfacial Thermal Conductance of TMDC Materials

|  | Thermal conductivity (W/mK) | | | Interfacial thermal conductance (MW/m²K) |
|---|---|---|---|---|
|  | Supported | Suspended (300K) | Suspended (500K) |  |
| 1L MoS$_2$ | 55±20 | 84±17 | 66±16 | 0.44±0.07[a] |
| 1L MoSe$_2$ | 24±11 | 59±18 |  | 0.09±0.03[b] |
| 2L MoS$_2$ | 35±7 | 77±25 | 29±10 | 0.74±0.05[a] |
| 2L MoSe$_2$ | 17±4 | 42±13 |  | 0.13±0.03[b] |

[a] Sample on Au. [b] Sample on SiO$_2$.

Finally, we use these values to model the response of suspended membranes, and iteratively determine the value of $\kappa$. Table 2 summarizes the calculated room-temperature values of $\kappa$, $\kappa_s$, and $g$ for the four sample types. The MoS$_2$ samples were prepared on the gold substrate and the MoSe$_2$ samples were prepared on the SiO$_2$ substrate. Since gold acts as a thermal sink, it results in a significantly higher interfacial thermal conductance. The measured value of $\kappa$ for suspended 1L MoS$_2$ (84±17 W/mK) is in good agreement with a recent theoretical prediction (83 W/mK).[38]

This value decreases at high temperature, as expected due to anharmonic phonon-phonon scattering. The thermal conductivity of MoSe$_2$ is somewhat smaller (59±18 W/mK). In both materials, the thermal conductivity of the 2L samples is smaller than in monolayers. This trend of decreasing thermal conductivity is also seen in graphene,[51] and can be attributed to a greater phase space for Umklapp phonon scattering[52] in thicker samples. All of these thin layers, in which there are no grain boundaries, have thermal conductivity much higher than that measured for bulk materials, which is of order 1 W/mK.[49,50]

The measured value of $\kappa$ for 1L MoS$_2$ is larger than the previously reported value of 34.5±4 W/mK for a 1.2 µm diameter membrane.[34] We find that most of the difference can be attributed to three factors. First, the prior works find the shift rates of 0.013 cm$^{-1}$/K, whereas we measure 0.020 cm$^{-1}$/K. The origin of this discrepancy is unclear: one possibility is that different-sized membranes may respond differently due to strain effects, but the A$_{1g}$ mode is largely insensitive to strain,[26] and reasonable values of thermal expansion coefficient lead to very small strain-induced shifts. Second, the prior work assumed optical absorption of 9.1%, whereas this work uses a value of 5.8%, based on measured dielectric functions as discussed above. Third, the prior work assumed interfacial thermal conductivity of $g$=10-50 MW/cm$^2$·K, over which range the derived value of $\kappa$ is largely insensitive to $g$, because the flake is well thermally grounded to the substrate. However, our measured values are substantially lower: 0.44 for 1L MoS$_2$ on gold and ~0.1 for 1L MoSe$_2$ on SiO$_2$. With g values in this range, the thermal decay length $\sqrt{\dfrac{\kappa_s t}{g}}$ is ~ 0.3-0.6 µm, such that the effective size of the suspended flake can be substantially larger than the hole diameter, particularly for diameters of order 1 µm. When previous data for shift rate vs. applied power for samples on Si$_3$N$_4$ is analyzed using our measured shift rate, $\kappa_s$, and g, we obtain a value of κ=84 W/mK, identical (within uncertainty) to the value obtained in this work. This result highlights the importance of accurate measurements of all experimental parameters for determination of thermal conductivity using the optothermal Raman method. Finally, we also note that the value of κ for 2L MoS$_2$ is comparable to experimental results on few-layer CVD MoS$_2$ (52 W/mK, with copper as substrate).[33]

The two published results of $g$ value for supported 1L MoS$_2$[34,54] have large discrepancy. In order to verify the value, and to check the validity of our analysis, we did a new type of experiment on the large size of the suspended membranes, as depicted in Fig. 4a. Using 1L MoS$_2$ on a 5μm diameter hole, the Raman laser (11μW absorbed power) was moved to a series of points away from the center of the membrane, with $0 < r_i < 1.75$ μm, and the temperature rise ΔT measured using the calibrated value from Table 1. The solid points in Fig. 4b represent the measured ΔT($r_i$), with 5 independent measurements at each position. Next, the expected value of ΔT($r_i$) is calculated by solving Eqs. 2 and 3 using finite element analysis (COMSOL). In the simulation, the interfacial thermal conductance $g$ is input and the value of κ is adjusted to fit the measured ΔT($r_i$=0). The simulation results for values of $g$ ranging from 0.1 MW/m$^2$K to 100 MW/m$^2$K are shown as solid lines in Fig. 4b. The experimental data closely follows the simulation for the previously calculated value of $g$ (0.44 MW/m$^2$K), and is clearly inconsistent with values above ~ 1 MW/m$^2$K. Although the value of g may very dependent on the details of a given sample (e.g. atomic-level cleanliness of the interface), these results conclusively demonstrate that the larger values of $g$ cannot be assumed (as in previous work[34]) without careful experimental verification. In addition, this method provides an independent check on measurements of $g$, a critical parameter for heat dissipation in electronic devices.

**CONCLUSION**

We have used a refined version of the optothermal Raman technique to study thermal transport 1L and 2L MoS$_2$, and the first such measurements for 1L and 2L MoSe$_2$. This work addresses several important issues in the measurement of thermal conductivity of 2D materials using Raman spectroscopy, in particular calibration of optical absorption and the role of thermal coupling to the substrate. We derive thermal conductivity values in good agreement with recent predictions for 1L MoS$_2$,[38] and find that the interfacial thermal conductance to the substrate is much smaller than previously assumed. The model is validated by studying the membrane response when heated away from the center. These results demonstrate more robust measurements of thermal transport in 2D materials, understanding of which is necessary for device modeling and other applications.

## EXPERIMENTAL SECTION

Samples were first exfoliated from bulk crystals on a bare Si wafer coated with polymethyl methacrylate (PMMA), and a thin release layer of polyvinyl alcohol (PVA) under the PMMA. The PMMA thickness of 280 nm provides good contrast for 1L and 2L flakes under an optical microscope. Each flake's thickness was determined by Atomic Force Microscopy (AFM) (Figure 1a) and Raman spectroscopy (Figure 1d). The PVA layer was dissolved in deionized water, leaving the PMMA-$MoS_2$ membrane floating on the water surface. The PMMA-$MoS_2$ ($MoSe_2$) membrane was then inverted and transferred[42] onto a gold-coated $SiO_2$ (280nm)/Si substrate with 2.5~5.0μm diameter holes etched to a depth of 1μm. The circular shape of the hole enables isotropic analysis of the radial thermal transport in the suspended flakes, and the comparatively large diameter ensures that the laser spot size is much smaller than the suspended area. The 10nm gold layer improves the interfacial thermal conductance between the supported flake and the substrate. The PMMA was then removed by annealing the sample under Ar/$H_2$ (21sccm/7sccm) flow at 350°C. A micro Raman spectrometer (RENISHAW InVia Raman Microscope system) was used to measure Raman peak shifts with temperature and laser power change. In order to obtain the relationship between Raman peak shift and temperature rise, samples were heated uniformly from 300K to 500K on a temperature-controlled heating platform (Linkam Stage THMS600). In order to prevent additional laser heating, Raman measurements were taken under low power (100 μW). Both the temperature dependent and the power dependent measurements were conducted with laser focused on the supported and the suspended flake areas.

*Conflict of Interest*: The authors declare no competing financial interest.

# FIGURES

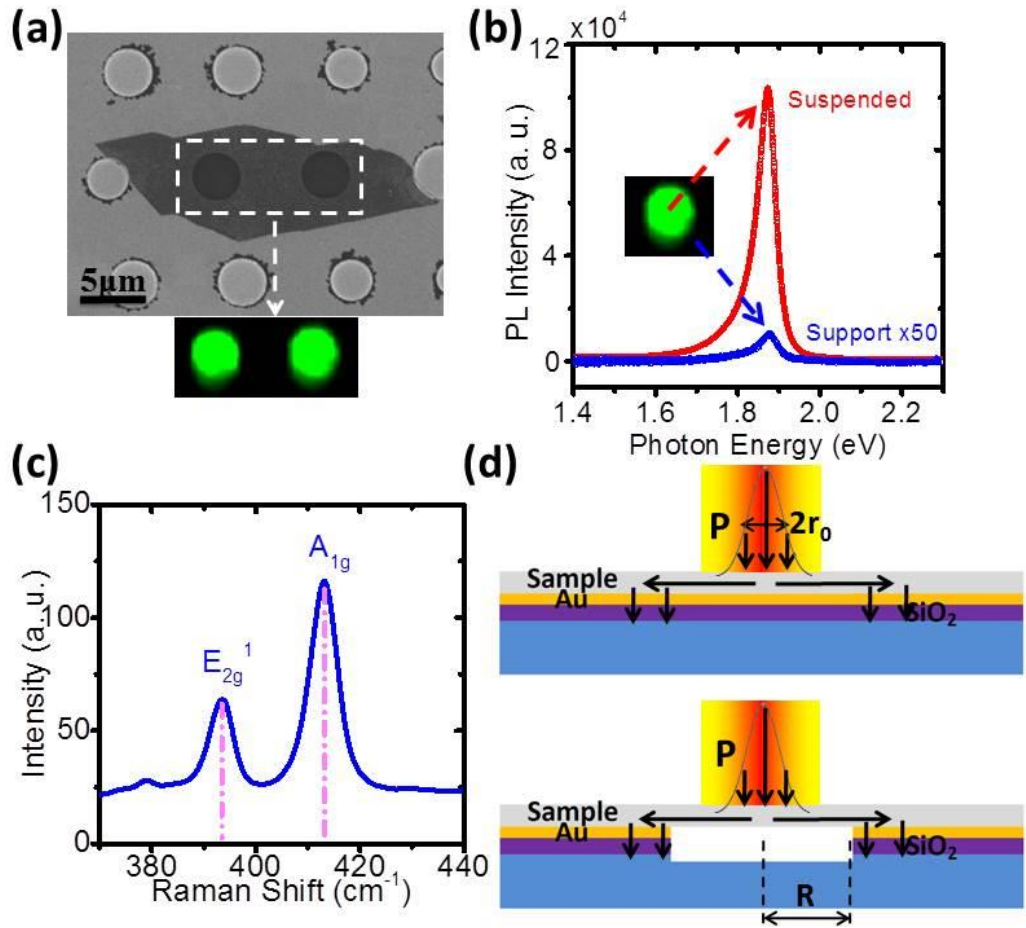

**Figure 1.** (a) The scanning electron microscopy image of the suspended 1L MoS$_2$ on the 4μm holes and its photoluminescence peak mapping. (b) Photoluminescence of the supported and suspended areas specifically. (c) The Raman spectra of 1L MoS$_2$. (d) Schematics of the experimental setup for the supported flake and the suspended flake.

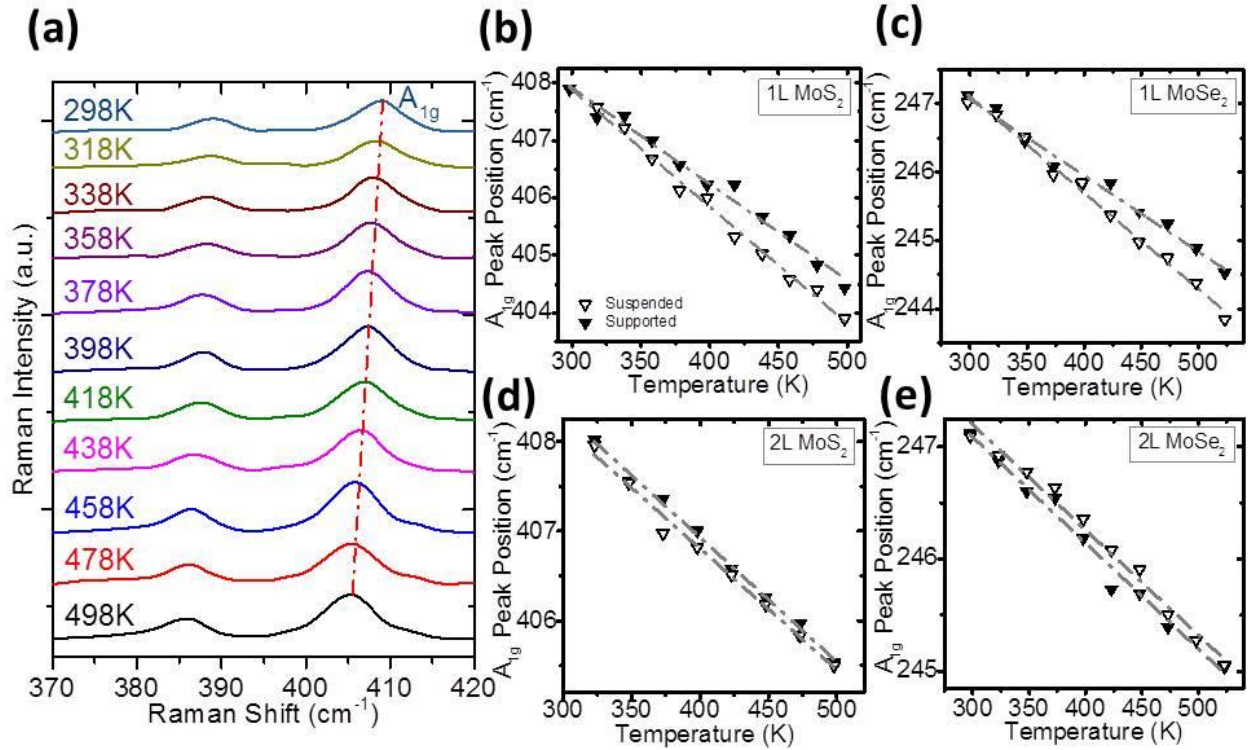

**Figure 2.** (a) Raman spectra of suspended 1L MoS$_2$ recorded at different temperatures. The temperature dependent A$_{1g}$ Raman peak shift measured on the supported and suspended 1L MoS$_2$ (b), 1L MoSe$_2$ (c), 2L MoS$_2$ (d), and 2L MoSe$_2$ (e).

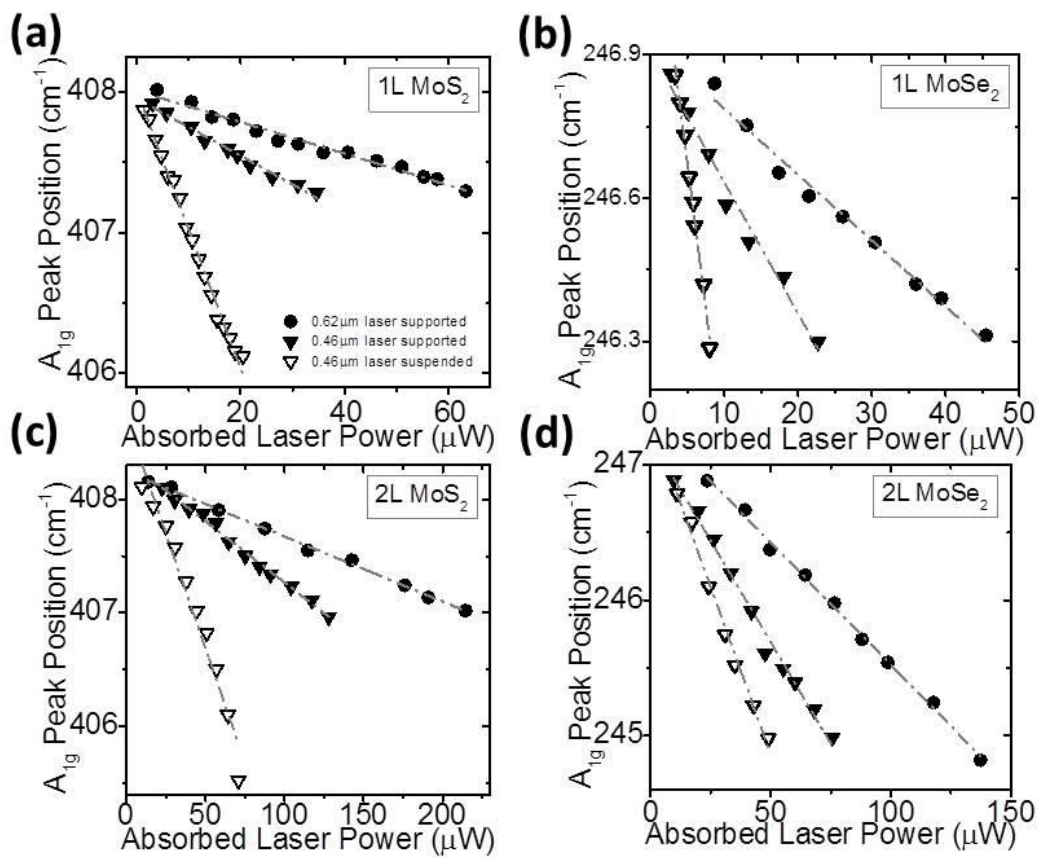

**Figure 3.** The power dependent $A_{1g}$ Raman peak shift measured using different laser spot sizes, on the supported and suspended 1L $MoS_2$ (a), 1L $MoSe_2$ (b), 2L $MoS_2$ (c), and 2L $MoSe_2$ (d).

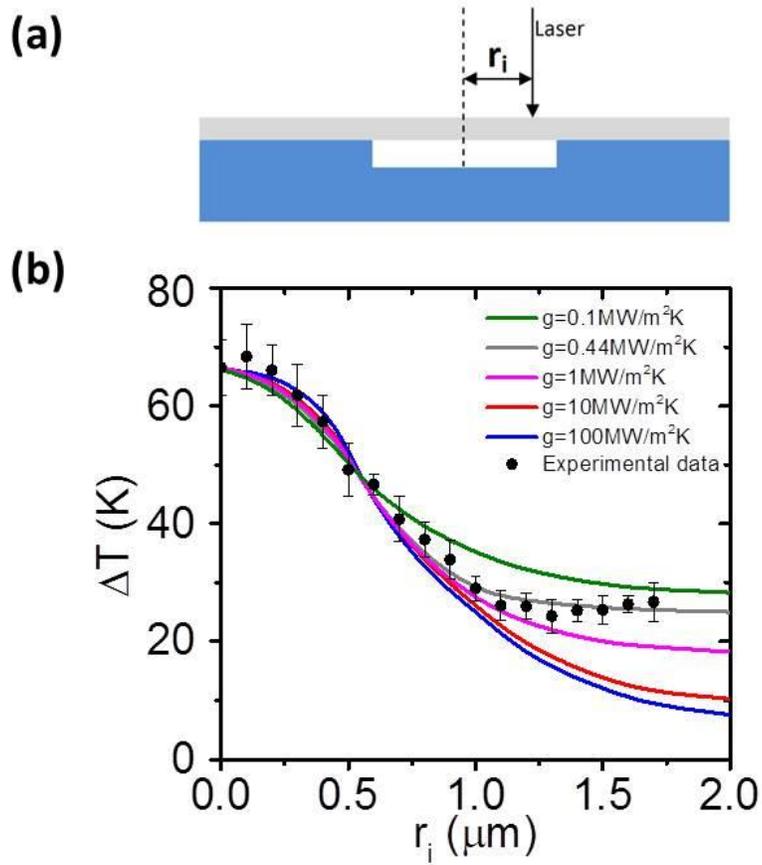

**Figure 4.** (a) The schematic of the Raman measurement at different sample positions. (b) The experimental and simulated position-dependent Raman curves on the suspended 1L $MoS_2$.

# AUTHOR INFORMATION

**Corresponding Author**

*jh2228@columbia.edu

**Author Contributions**

The manuscript was written through contributions of all authors. All authors have given approval to the final version of the manuscript.

**Acknowledgements**

This work was supported by AFOSR under award FA9550-14-1-0268. G.H.L. was supported by Basic Science Research Program (NRF-2014R1A1A1004632) through the National Research Foundation (NRF) funded by the Korean government Ministry of Science, ICT and Future and in part by the Yonsei University Future-leading Research Initiative of 2014.